\documentstyle[aps,epsfig]{revtex}

\newcommand{\be}{\begin{equation}}
\newcommand{\ee}{\end{equation}}
\newcommand{\bea}{\begin{eqnarray}}
\newcommand{\eea}{\end{eqnarray}}

\def\la{\mathrel{\mathpalette\fun <}}

\def\fun#1#2{\lower3.6pt\vbox{\baselineskip0pt\lineskip.9pt
\ialign{$\mathsurround=0pt#1\hfil##\hfil$\crcr#2\crcr\sim\crcr}}}

\begin{document}

\title{Determination of hadronic partial widths for
scalar-isoscalar resonances  $f_0(980)$, $f_0(1300)$, $f_0(1500)$,
$f_0(1750)$ and the broad state $f_0(1530\; ^{+\; 90}_{-250})$}

\author{V. V. Anisovich, V. A. Nikonov, A. V. Sarantsev}
\date{\today}
\maketitle

\begin{abstract}
In the article of V.V. Anisovich et al., Yad. Fiz. {\bf 63}, 1489
(2000), the $K$-matrix solutions for the wave $IJ^{PC}=00^{++}$ were
obtained in the mass region 450--1900 MeV where four resonances
$f_0(980)$, $f_0(1300)$, $f_0(1500)$, $f_0(1750)$ and the broad state
$f_0(1530\; ^{+\; 90}_{-250})$ are located.
Based on these solutions,
we determine partial widths for scalar-isoscalar states decaying
into the channels $\pi\pi$, $K\bar K$, $\eta\eta$, $\eta\eta'$,
$\pi\pi\pi\pi$ and corresponding decay couplings.
\end{abstract}

\section{Introduction}

In \cite{YF} the combined $K$-matrix analysis was performed for the
meson partial waves $IJ^{PC}=00^{++},10^{++},02^{++},12^{++}$ on the
basis of GAMS data on $\pi^-p\to\pi^0\pi^0 n$, $\eta\eta n$, $\eta\eta'
n$ \cite{GAMS}, BNL data on $\pi^-p\to K\bar K n$ \cite{BNL} and
Crystal Barrel data on $p\bar p (at \; rest)\to\pi^0\pi^0\pi^0$,
$\pi^0\eta\eta$, $\pi^0\pi^0\eta$ \cite{CB}. The positions of
amplitude poles (physical resonances) were determined together with the
positions of the $K$-matrix poles (bare states) and  bare-state
couplings to the two-meson channels. The nonet classification of the
bare $q\bar q$ states was suggested, and the possibilities for the
location of the lightest scalar glueball in the mass region 1200--1700
MeV were discussed.

The $K$-matrix technique has an advantage of taking account of the
unitarity condition constraints and, in this way, of a
correct incorporation of  threshold singulatities
into the scattering amplitude.
Since the search for resonances is always related to the
investigation of analytical structure of the amplitude in the
complex-mass plane by using data at real masses,
it is important to perform analytical continuation of the amplitude into
the lower half-plane of the complex mass, with correctly taken
singularities on the real axis.

However, the $K$-matrix amplitude does not include
  resonance parameters in an explicit
form, so  additional calculations  are needed  to determine
masses and couplings of real resonances.
In paper \cite{YF} the pole positions have been found
for the considered partial wave amplitudes  (i.e. masses and
total widths of resonances were found), but  more complicated
calculation of couplings has not been done yet. The decay coupling
constants are to be determined as residues of the pole singularities
of the multi-channel amplitude. In the present paper we overcome this
deficiency and calculate  coupling constants to the channels
$\pi\pi$, $K\bar K$, $\eta\eta$, $\eta\eta'$ and $\pi\pi\pi\pi$
for the resonances $f_0(980)$,
$f_0(1300)$, $f_0(1500)$, $f_0(1750)$, $f_0(1530\; ^{+\; 90}_{-250})$.
This procedure provided us with partial widths for the decays of these
resonances.
The choice of the scalar-isoscalar sector for
primary study follows from the interest in pursuing
the destiny of the lightest
scalar glueball after its  mixing with neighbouring states;
one needs a
knowledge of the couplings $f_0 \to \pi\pi, K\bar K, \eta\eta,
\eta\eta'$ for all resonances over the mass region 1000--1800 MeV.

The paper is organized as follows.

In Section 2 the coupling
constants are presented for the decays $f_0\to \pi\pi, K\bar K,
\eta\eta,\eta\eta', \pi\pi\pi\pi$, and partial widths for the mesons
 $f_0(1300)$, $f_0(1500)$, $f_0(1750)$ and $f_0(1530^{+ 90}_{-250})$
are determined.

In Section 3 the resonance $f_0(980)$ is considered
in detail: the results of the analysis \cite{YF} tell us
that standard formulae for the description of resonances, such as
Breit-Wigner or Flatt\'e ones, in the case of  $f_0(980)$
are unable to give simultaneously the values of the decay coupling
constants and the position of the amplitude pole.  We
suggest an alternative form of resonance amplitude for $f_0(980)$ in
which the important role is played by the prompt transition $\pi\pi\to
K\bar K$. In this Section another low-mass state, namely, the
$\sigma$-meson, is also discussed.

The results are summarized in the Conclusion.

\section{Decay couplings and partial decay widths}

We determine the coupling constants and partial decay widths using
the following procedure.
The $00^{++}$-amplitude for the transition $a\to b$,
\be
A_{a\to b}(s)\; , \qquad a,b=
\pi\pi,K\bar K, \eta\eta, \eta\eta', \pi\pi\pi\pi
\label{1}
\ee
is considered as a function of the invariant energy squared $s$
in the complex-$s$ plane near the pole related to the
resonance $n$. In the vicinity of the pole the amplitude reads:
\be
A_{a\to b}(s)= \frac{g^{(n)}_a g^{(n)}_b}{\mu^2_n -s}
\; e^{i\theta ^{(n)}_{ab}} + B_{ab}\; .
\label{2}
\ee
Here $\mu_n$ is the resonance complex mass $\mu_n =M_n -i\Gamma_n/2\;$;
$g^{(n)}_a$ and $ g^{(n)}_b$ are the couplings for
the transitions $f_0 \to a$ and $f_0 \to b$.
The factor $ \exp{(i\theta ^{(n)}_{ab}}) $ is due to a background
contribution which can
be the non-resonance terms or tails of neighbouring
resonances. We also write down in (\ref{2}) the non-pole background
term $B_{ab}$.

The partial width for the decay  $f_0 \to a$ is determined as a
product of the coupling  constant
squared, $g^{(n)\; 2}_a$, and  phase space,
$\rho_a (s)$, averaged over resonance density:
\be
\Gamma_a (n)= C_n \int \limits_{s>s_{th}} \frac{ds}{\pi} \,
\frac{g^{(n)\; 2}_a \;\rho_a (s)}{(Re \;\mu^2_n -s)^2+
({\rm Im} \,\mu^2_n)^2}
\ .
\label{3}
\ee
Following \cite{YF},  we write down the phase space factor as follows:
\be
\rho_a (s)= \frac{2k_a}{\sqrt s}\ ,
\label{4}
\ee
where $k_a$ is the relative momentum of mesons in the decay channel (for
example, for the $\pi\pi$ channel
$\rho_{\pi\pi} (s)=\sqrt{(s-4m^2_{\pi})/s}$ ). For the $\pi\pi\pi\pi$
channel the phase space factor was chosen in \cite{YF}
to be the same as for the
two-$\rho$-meson state at $s< 1$ GeV$^2$ or
be equal to $1$ at $s\geq 1$ GeV$^2$.
The integration over $s$ in (\ref{3}) is carried
out in the region above the
$a$-channel threshold, $s>s_{th}$ (for the $\pi\pi$ channel it is
$s>4m^2_{\pi}$). The resonance density factor,
$\left ((Re \;\mu^2_n -s)^2+({\rm Im} \,\mu^2_n)^2\right )^{-1}$
guarantees rapid convergence of the integral (\ref{3}).
The normalization constant $C_n$ is determined by the requirement that
the sum
of all hadronic partial widths is equal to the total width of the
resonance:
\be
\Gamma (n) =\sum \limits_{a} \Gamma_a (n)\ .
\label{5}
\ee

In paper \cite{YF} three solutions for the wave $IJ^{PC}=00^{++}$ have
been found; they are labelled as I, II-1 and II-2 (see Tables 4 and 5
in \cite{YF}). In practice Solutions II-1 and II-2 give the same
physical parameters of resonances, though they differ from
parameters found for the $K$-matrix elements. In particular, in
Solution II-2 the state $f^{bare}_0(1600)$ may be identified as a
gluonium, for the decay couplings satisfy all the requirements
inherent to glueball state; in Solution II-1 such
a state is $f^{bare}_0(1230)$. For  Solution I, the same bare state,
$f^{bare}_0(1230)$, should be considered as a gluonium.

In Table 1 we show the values of partial widths
for the resonances
$f_0(980)$, $f_0(1300)$, $f_0(1500)$, $f_0(1750)$ and broad state
$f_0(1530\; ^{+\; 90}_{-250})$.
Partial widths for
$f_0(1300)$, $f_0(1500)$, $f_0(1750)$,
$f_0(1530\; ^{+\; 90}_{-250})$
are calculated within standard formulae for the
Breit-Wigner resonances (\ref{3}), (\ref{4}) and (\ref{5}).
The resonance $f_0(980)$ being located near the strong $K\bar K$
threshold needs a special consideration that is presented
below.

The decay
coupling constants squared, $g^{(n)\; 2}_a$, are shown in Table 2
for $a=\pi\pi, K\bar K, \eta\eta, \eta\eta', \pi\pi\pi\pi$.
The couplings are determined with
the normalization of the amplitude used in \cite{YF}:
for example, we write
the $\pi\pi$ scattering amplitude
(\ref{2}) as $A_{\pi\pi\to \pi\pi}(s)= (
\eta^0_0 \exp{(2i\delta^0_0)}-1)/2i\rho_{\pi\pi}(s)$, where $\eta^0_0$
and $\delta^0_0$ are
the inelasticity parameter and phase shift for the
$00^{++}$ $\pi\pi$-wave. The coupling constants $g^{(n)\; }_a$ are
found by calculating
the residues of the amplitudes $\pi\pi\to \pi\pi$,
$K\bar K$, $\eta\eta$, $\eta\eta'$, $\pi\pi\pi\pi$.
Also we check
the factorization property for the pole terms by
calculating residues for other reactions,
such as $K\bar K \to K\bar K $.

The position of poles
in the complex-$M$ plane ($M\equiv \sqrt s$) is illustrated by Fig. 1.
The complex-$M$ area, where the $K$-matrix fit \cite{YF} may  reliably
reproduce analyticity of the  amplitude, is
inside a semi-circle depicted by dashed line.
The poles which are a subject of the $K$-matrix analysis and correspond
to  $f_0(980)$,
$f_0(1300)$, $f_0(1500)$, $f_0(1750)$,
$f_0(1530\; ^{+\; 90}_{-250})$ are located on the 3rd, 4th, 5th and 6th
sheets of the complex-$M$ plane. The resonance $f_0(980)$ is located
near the strong $K\bar K$ threshold, therefore two poles
are related to $f_0(980)$: the nearest one is on the 3rd sheet
($M\simeq 1014-i39$ MeV) and a remote pole
 on the 4th sheet ($M\simeq 936-i238$ MeV). Coupling
constants for  $f_0(980)$ are determined as residues of the nearest
pole (on the 3rd sheet). The $\eta \eta'$ threshold is weak for
$f_0(1500)$, and because of that positions of poles on the 5th and
6th sheets are  practically the same (note that couplings related to
these poles nearly coincide).

The $K$-matrix fit \cite{YF} has been carried out in a broad
mass in\-ter\-val, $450\, {\rm MeV}\,\le M \le 1900$ MeV. This
very fact allows us to believe that we deal with successfully
reconstructed analytical amplitude which goes rather deeply into
lower half-plane $M$, and this area is restricted  by  dashed
line in Fig. 1. It is of crucial
importance that the pole of the broad state
$f_0(1530^{+90}_{-250})$ is inside this area, for this broad state
plays a key role in a mixing of $q\bar q$-mesons with the lightest
glueball, see \cite{AAS-Zeit,ABS-PR} for details.

\section{The low-mass mesons: $f_0(980)$ and $\sigma$}

The two low-mass mesons, $f_0(980)$ and $\sigma$-meson, need special
consideration and comments.

The analysis \cite{YF} shows us that  $f_0(980)$  cannot be
described either by standard Breit-Wigner formula or its modification
for the case of the nearly located strong $K\bar K$ threshold, that
is, Flatt\'e's formula \cite{Flatte}.  Here we suggest another
resonance formula for $f_0(980)$ which agrees with the results of
\cite{YF}.

In the compilation of Particle Data Group \cite{PDG},
the $\sigma$-meson is denoted as $f_0(400-1200)$ that
reflects a cumulative result obtained in a number of papers where the
mass of $\sigma$ was found in this region or even
higher.  However, the analysis \cite{YF} definitely demonstrates the
absence of poles in the $00^{++}$-amplitude  at
$(600 \le Re\; M \le 1200)$ MeV,
with an exception of poles for $f_0(980)$
 -- we will discuss the situation with $\sigma$-meson in this
Section later on.

\subsection{ Description of $f_0(980)$ }

For $f_0(980)$, the $K$-matrix fit \cite{YF} gives us the position of
the pole and coupling constant values, see Tables 1 and 2. These
parameters are sufficient to reconstruct the Breit--Wigner resonance
amplitude. However, in case of $f_0(980)$ there exists a strong $K\bar
K$ threshold near the pole, so the resonance term in the amplitude
(\ref{2}) should be suggested not as the Breit-Wigner pole but in a
more complicated form.  For the
$\pi\pi\to \pi\pi$ and $K\bar K\to K\bar K$
transitions near $f_0(980)$, the following resonance amplitudes can be
written instead of the Breit-Wigner pole term
$R^{(ab)}_n =g^{(n)}_a g^{(n)}_b/(\mu^2_n -s)$ entering
equation (\ref{2}):
\bea
R^{(\pi\pi, \pi\pi)}_{f_0(980)}
=\left (G^2+i\frac{\sqrt{s-4m^2_K}}{m_0}\; F\right )\,\frac 1D\ , \qquad
R^{(K\bar K, K\bar K)}_{f_0(980)}
=\left (G^2_{K\bar K}+i\,F\right )\frac 1D\ ,
\label{star}
\eea
where
\bea
&&F=2GG_{K\bar K}f+f^2(m^2_0-s)\ ,\nonumber \\
&&D=m^2_0-s-iG^2-i
\frac{\sqrt{s-4m^2_K}}{m_0}
\left (G^2_{K\bar K}+i\,F\right )\ .
\eea
Here  $m_0$ is the input mass of $f_0(980)$, $G$ and $G_{K\bar K}$
are coupling constants to pion channels
($\pi\pi+\pi\pi\pi\pi$) and $K\bar K$. The dimensionless constant
$f$ stands for the prompt transition $ K\bar K\to\pi\pi$:
the value $f/m_0$ is the "transition length" which is analogous
to the scattering length of the low-energy hadronic interaction.
The constants
$m_0$, $G$, $G_{K\bar K}$, $f$
are parameters which are to be chosen to reproduce
the $f_0(980)$ charactristics (position of pole $s\simeq (1.02
-i0.08)$ GeV$^2$ and  couplings to the channels $\pi\pi$
and $K\bar K$, $g^2_{\pi\pi}\simeq 0.076$ GeV$^2$ and
$g^2_{K\bar K}\simeq 0.184$ GeV$^2$, see Tables 1 and 2).

The $\pi\pi$ scattering amplitude in the $f_0(980)$ region is now
defined as
\be
A_{\pi\pi\to\pi\pi}=
 e^{i\theta}\; R^{(\pi\pi, \pi\pi)}_{f_0(980)}+
e^{i\frac{\theta }{2} }
\sin\frac{\theta }{2}\ .
\label{pi-pi}
\ee
This formula may be compared with
equation (\ref{2}): the background term in (\ref{pi-pi}) is fixed by
the requirement that $\pi\pi$ scattering
amplitude below $K\bar K$ threshold has the form
$\exp{(i\delta)}\sin{\delta}$.

At $f\to 0$ the resonance
amplitudes (\ref{star}) turn into Flatt\'e's  formula
\cite{Flatte}, which is used rather often for the description of
$f_0(980)$ .  Still, it happened that position of the pole (complex
mass value) as well as the amplitude residue in the pole, which have
been determined in \cite{YF} and shown in Tables 1, 2, do not obey the
Flatt\'e formula but require $f\ne 0$.

We obtained two sets of parameters, with sufficiently correct
values of the $f_0(980)$ pole position and couplings. They are
equal (in GeV units) to:
\bea
&&Solution \,A: \quad m_0=1.000\ ,\,f=0.516\ ,\,G=0.386\ ,\,
G_{K\bar K}=0.447\ ,\nonumber \\
&&Solution \,B: \quad m_0=0.952\ ,\,f=-0.478\ ,\,G=0.257\ ,\,
G_{K\bar K}=0.388\ .
\label{AB}
\eea
The above parameters  provide us with a reasonable description of
the $\pi\pi$ scattering amplitude. The phase shift $\delta^0_0$ and
inelastisity parameter $\eta^0_0$ are shown in Fig. 2; the angle
$\theta$ for the background term in Solutions A and B,
determined as
\be
\theta = \theta_1 +(\frac{\sqrt{s}}{m_0} -1)\theta_2\; ,
\ee
is numerically equal to
\bea
&&Solution \,A: \quad
 \theta_1=189^\circ, \;\; \theta_2=146^\circ \; ,
\nonumber \\
&&Solution \,B: \quad
 \theta_1=147^\circ, \;\; \theta_2=57^\circ
\ .
\label{theta}
\eea
Solutions A and B give significantly different predictions for
$\eta^0_0$; however, the existing data \cite{ochs,kmAS} do not allow
us to discriminate between them.

Partial widths of $f_0(980)$ are calculated with
the expression similar to (3) --- with the replacement of the integrand
denominator  as follows:
\be
({\rm Re}\,\mu_n^2-s)^2+ ({\rm Im}\,\mu_n^2)^2 \to |D|^2.
\ee
For both sets of parameters
(\ref{AB}) the calculated partial widths are close to
each other. For example, using Solution II-2 and
the $A$-set of parameters  we have
$\Gamma_{\pi\pi}=62$ MeV, $\Gamma_{K\bar K}=14$ MeV, while for solution
II-2 and the $B$-set of the
parameters one has $\Gamma_{\pi\pi}=66$ MeV, $\Gamma_{K\bar K}=10$
MeV. The values of partial widths for $f_0(980)$ averaged
over Solutions $A$ and $B$ are presented  in Table 1.

The total hadron width of $f_0(980)$ is defined in the
same way as for
the other $f_0$-mesons, namely, by using the position of pole in the
complex-$M$ plane: the imaginary part of the mass is equal to a
half-width of the resonance. For the Breit--Wigner resonance this
definition is in accordance with what is observed from resonance
spectrum (provided there is no interference with the background). If
the resonance is located in the vicinity of a strong threshold, the
observed resonance width can differ significantly from what is given by
the pole position. In Fig. 3 one can see the magnitudes $|R^{(\pi\pi,
\pi\pi)}_{f_0(980)}|^2\rho_\pi (s)$, $|R^{(K\bar K, K\bar
K)}_{f_0(980)}|^2\rho_K (s)$ and $g_{\pi\pi}^2 \Gamma_{tot}/|D(s)|^2$
for the parameter sets A and B: the
width of peaks does
vary, being less than the value determined
by the complex mass of the resonance.

The only objective characteristic of the total hadron width is the
position of pole in the complex-$M$ plane, due to this
reason we employ such a
definition of total hadron width. Multiple variations of total
width in the compilation \cite{PDG} are just due to the absence of
a proper definition of $\Gamma_{tot}$ for resonances near the strong
threshold.

\subsection{The light $\sigma$-meson }

The light $\sigma$-meson reveals
itself as a pole on the 2nd sheet: it is
shown in Fig. 1 at $M=(431-i325)$ MeV (or, in terms of $s$
which is more appropriate variable for light particles, at
$s=(4-i14)m_\pi^2$ ) that corresponds to
the magnitude obtained in \cite{sigma}. Although
this pole does not appear in the area of complex $M$, where the
$K$-matrix fit \cite{YF} reconstructs the amplitude rather reliably, it
still deserves detailed comments.

The situation with $\sigma$-meson is as follows.
The $K$-matrix representation allows us to reconstruct correctly the
analytical structure of the
partial amplitude in the physical region, at
$s\ge 4m_\pi^2$ by taking account of the threshold and pole
singularities. The singularities related to forces
(or left singularities,
at $s\le 0$), are not included directly into the $K$-matrix machinery.
This does not allow us to be quite sure about the results of the
$K$-matrix approach at $s\la 4m_\pi^2$. Concerning the low-mass region,
$s > 4m_\pi^2$,
an important result of the $K$-matrix fit \cite{YF} is the absence
of the pole singularity in the $00^{++}$ amplitude at 500--800 MeV. Here
the $\pi\pi$-scattering phase $\delta^0_0$ increases smoothly reaching
$90^\circ$ at 800--900 MeV. A straightforward explanation of such a
behaviour of $\delta^0_0$ could consist in the existence of a broad
resonance, with the mass about 600--900 MeV and width $\sim$800 MeV
(for example, see discussion in \cite{Montanet,Penn} and references
therein). However, as was stressed above,
the $K$-matrix amplitude does not contain pole
singularities at $500 \le Re\,M \le 900$ MeV: the $K$-matrix amplitude
has a low-mass pole only, which is located near the $\pi\pi$ threshold
or below it. In \cite{YF} the presence of the pole near the
$\pi\pi$ threshold  was not emphasized, since
the $K$-matrix solution does not guarantee a
reliable reconstruction of
the amplitude at $s\sim 4m_\pi^2$.
In \cite{sigma}, in order
to restore analytical structure at  $s\sim 4m_\pi^2$,
the left-hand-side singularities were
accounted for
on the basis of the dispersion relation $N/D$-method.
The $\pi\pi$ scattering $N/D$-amplitude was
represented at $M\le 900$  MeV
being sewed with the $K$-matrix solution \cite{YF}
at $450 \le M \le 900$ MeV.
The $N/D$-amplitude reconstructed in this way has a pole near
the $\pi\pi$ threshold, thus proving that qualitatively the results
of \cite{YF} are also valid for the region $s\sim 4m_\pi^2$.
The pole of the $N/D$-amplitude \cite{sigma} is shown in Fig. 1.

It is worth  mentioning that the low-mass location of the
$\sigma$-meson pole was also obtained in a set of papers, where the
low-energy $\pi\pi$ amplitude has been investigated by taking into
account the left-hand cut as a set of meson exchanges.
These papers include:  (i) dispersion relation approach, $s \simeq
(0.2-i22.5)m_\pi^2 $ \cite{Basdevant}, (ii) meson exchange models, $s
\simeq (3.0-i17.8)m_\pi^2 $ \cite{Zinn}, $s \simeq (0.5-i13.2)m_\pi^2 $
\cite{Bugg}, $s \simeq (2.9-i11.8)m_\pi^2 $ \cite{Speth}, (iii) linear
$\sigma$-model, $s \simeq (2.0-i15.5)m_\pi^2 $ \cite{Achasov}.
At the same time, in \cite{800P,800E,800A,800Ishida,Roos,Loch} the pole
position was found in the region of higher $s$, at $s> 7
m_\pi^2$, that reflects the ambiguities of approaches which
treat the left-hand cut as a known quantity.

As to finding out the location of $\sigma$-meson on the basis of the
available experimental data, one should make general remark. Since the
width of the  $\sigma$-meson is rather large, it is necessary to fit to
data in the energy interval  which is much larger than a  total
width of the $\sigma$-meson. For example, to speak about $\sigma$-meson
with a mass $M_\sigma \sim 900$ MeV and half-width $\Gamma/2 \simeq
400$ MeV, one should fit to data in the interval $300\;{\rm MeV} \la M
\la 1400$ MeV and at the same time to take a correct account of the
nearest singularities, which are poles corresponding to $f_0(980)$,
$f_0(1300)$ and presumably $f_0(1500)$ as well as threshold
singularities $\pi\pi$, $K\bar K$, $\eta, \eta'$ and $\pi\pi\pi\pi$.
Concerning $\pi\pi\pi\pi$, one should have in mind that the
contribution of this channel is significant starting from 1300 MeV, so
this channel is absolutely necessary. Such demands  towards the
fit of experimental data have been fulfilled in no paper under
discussion, with an exception for \cite{YF,kmAS}.

\section{Conclusion}

We have obtained partial decay widths for five scalar-isoscalar
states
$f_0(980)$, $f_0(1300)$,
$f_0(1500)$, $f_0(1750)$, $f_0(1530^{+90}_{-250})$  by calculating the
decay couplings as residues of pole singularities in the $K$-matrix
amplitude \cite{YF}: positions of poles in the complex-$M$ plane
are shown in Fig. 1. The pole which corresponds to the light
$\sigma$-meson is also shown in Fig. 1: it was not included into the
$K$-matrix calculation procedure directly, being close to the left-hand
cut; the discussion of its status can be found in
\cite{sigma} and references therein.

The results of our calculations of partial decay widths are
presented below (the magnitudes are given in MeV units):

\vspace{0.5cm}
\begin{tabular}{lcccccc}
Resonance &$\Gamma_{\pi\pi}$&$\Gamma_{K\bar K}$&
$\Gamma_{\eta\eta}$&$\Gamma_{\eta\eta'}$&$\Gamma_{\pi\pi\pi\pi}$&
$\Gamma_{tot}/2$\\

$f_0(980):$ &$64\pm 8$ & $12\pm 1$ & -- & -- & $4\pm 2$ & $40\pm 5$\\

$f_0(1300):$&$46\pm 12$ & $5\pm 3$ & $4\pm 2$  & -- & $171\pm 10$ &
$113\pm 10$\\

$f_0(1500):$&$37\pm 2$ & $7\pm 2$ & $4\pm 1$ & $0.2\pm 0.1$ &
$79\pm 6$ & $62\pm 3$\\

$f_0(1750):$&$74^{+15}_{-30}$ &  $11^{+17}_{-9}$ & $7\pm 1$ & $3\pm 1$ &
$91^{+30}_{-60}$ &$93^{+20}_{-40}$\\

$f_0(1530^{+90}_{-250}):$&$380\pm 15$ & $185\pm 10$ & $40\pm 5$ &
$1\pm 1$ & $560^{+260}_{-125}$ & $590^{+120}_{-200}$\\
\end{tabular}
\be
\;
\label{table}
\ee
The  values shown for partial widths as well as  decay coupling
constants
of Table 2 need some comments.

The comparison of the hadron decays $f_0(980) \to K\bar K $ and
$f_0(980) \to \pi\pi $ points to  a large $s\bar s$
component in $f_0(980)$.
The analysis of radiative decays $\phi (1020) \to \gamma f_0(980)$
and $ f_0(980) \to \gamma\gamma$ \cite{phi} shows also that the $s\bar
s$ component in $f_0(980)$ is large: with the $f_0(980)$ flavour
wave function written as $n\bar n \cos \varphi +s\bar s \sin \varphi$,
the radiative decay widths give either $\varphi \simeq -48^\circ$
or  $\varphi \simeq 86^\circ$ (solution with  negative
$\varphi $ is more preferable).
When the decay processes are switched off,
$f_0(980)$ transforms into $f^{bare}_0(720\pm 100)$,
with $\varphi_{bare} \simeq -70^\circ$
(corresponding pole trajectory in the complex-$M$ plane is shown in
Fig.  10 of \cite{YF}).
 We see that the decay processes and related  change
of the state do not diminish the $s\bar s$ component strongly.

An opposite situation takes place with $f_0 (1750)$. After switching
off the decay channels, this resonance transforms into
$f^{bare}_0(1810\pm 30)$ which is dominantly $s\bar s$:
$\varphi_{bare} \simeq 90^\circ$ for  Solution I and $\varphi_{bare}
\simeq - 60^\circ$ for Solution II. However, partial decay
widths of $f_0 (1750)$ (or decay coupling constants given in Table 2)
unambiguously prove that $s\bar s$ component in $f_0 (1750)$
decreased strongly due to a mixing with other states after the
 onset of the  decay processes.  It is
possible to guess that this $s\bar s$ component has flown into the
broad state $f_0(1530^{+90}_{-250})$: the ratio $\Gamma_{K\bar
K}/\Gamma_{\pi\pi} $ for $f_0(1530^{+90}_{-250})$ does not contradict
such an assumption.  Such a scenario looks rather intriguing, in
particular when taking account of the fact that the broad state
$f_0(1530^{+90}_{-250})$, according to \cite{YF}, is a descendant of a
pure glueball (see also \cite{AAS-Zeit,AS,A-UFN}).  However, the
 study of the mixing of  $q\bar q$-state with the glueball is beyond
the frame of this article; it will be investigated elsewhere.

For $f_0(980)$, the obtained magnitudes for the complex mass and
decay couplings $g^2_\pi$ and $g^2_K$, demonstrate a failure  of
the Flatt\'e formula. We suggest an alternative description of
$f_0(980)$ which explores, as an addition to the pole term, the
amplitude for the prompt transition $\pi\pi \to K\bar K$.

\section*{Acknowledgement}
The authors are indebted to
D.V. Bugg,  L.G. Dakhno and L. Montanet for illuminative discussions of
problems related to scalar-isoscalar resonances.
The work was supported by the RFBR grant 01-02-17861.

\newpage
\begin{table}
\caption{Partial widths of scalar-isoscalar resonances (in MeV units)
in hadronic channels $\pi\pi$, $K\bar K$, $\eta\eta$, $\eta\eta'$ and
$\pi\pi\pi\pi$ for different $K$-matrix solutions of Ref. [1].}
\begin{tabular}{l|ccccc|l|c}
\hline
 &$\pi\pi$&$K\bar K$&$\eta\eta$&$\eta\eta'$&$\pi\pi\pi\pi$&
pole position&solution\\
\hline
$f_0(980)$&71&13&--&--&6&1006 - i 45  &I\\
          &56&10&--&--&2&1020 - i 34  &II-1\\
          &64&12&--&--&3&1015 - i 39.5&II-2\\
\hline
$f_0(1300)$&71&12&7&--&170&1307 - i 130 &I\\
           &34&2 &2&--&171&1296 - i 104 &II-1\\
           &34&2 &2&--&171&1296 - i 104 &II-2\\
\hline
$f_0(1500)$&38&9&4&0.3&83&1495 - i 67 &I\\
           &37&6&4&0.1&83&1498 - i 60 &II-1\\
           &37&6&4&0.1&72&1495 - i 60 &II-2\\
\hline
$f_0(1750)$&45&28&6&4&29 &1781 - i 56 &I\\
           &88&2 &7&3&120&1817 - i 110&II-1\\
           &90&2 &7&3&123&1817 - i 113&II-2\\
\hline
$f_0(1530^{+90}_{-250})$&376&173&42&2&818&1609 - i 706&I\\
                        &377&196&39&1&434&1427 - i 526&II-1\\
                        &376&196&39&1&438&1430 - i 525&II-2\\
\hline
\end{tabular}
\end{table}

\begin{table}
\caption{Coupling constants squared (in GeV$^2$ units) of
scalar-isoscalar resonances to hadronic channels $\pi\pi$, $K\bar
K$, $\eta\eta$, $\eta\eta'$ and $\pi\pi\pi\pi$ for different $K$-matrix
solutions of Ref. [1].}
\begin{tabular}{l|ccccc|c}
\hline
 &$\pi\pi$&$K\bar K$&$\eta\eta$&$\eta\eta'$&$\pi\pi\pi\pi$&solution\\
\hline
$f_0(980)$&0.076 & 0.180 &  0.075 &  -- &   0.009 &I\\
          &0.076 & 0.186 &  0.072 &  -- &   0.004 &II-1\\
          &0.076 & 0.186 &  0.072 &  -- &   0.004 &II-2\\
\hline
$f_0(1300)$&0.050 &  0.015 &  0.012 &  -- & 0.124 &I\\
           &0.026 &  0.002 &  0.003 &  -- & 0.132 &II-1\\
           &0.026 &  0.002 &  0.003 &  -- & 0.132 &II-2\\
\hline
$f_0(1500)$&0.032 &  0.010 &   0.005 &  0.012 &  0.070 &I\\
          & 0.038 &  0.009 &   0.007 &  0.006 &  0.074 &II-1\\
          & 0.038 &  0.009 &   0.007 &  0.006 &  0.074 &II-2\\
\hline
$f_0(1750)$&0.039 &  0.029 &   0.007 &  0.030 & 0.025 &I\\
          & 0.086 &  0.003 &   0.009 &  0.028 & 0.117 &II-1\\
          & 0.086 &  0.003 &   0.009 &  0.028 & 0.117 &II-2\\
\hline
$f_0(1530^{+90}_{-250})$&0.329 & 0.229 &  0.061 &  0.022 & 0.764 &I\\
                        &0.304 & 0.271 &  0.062 &  0.016 & 0.382 &II-1\\
                        &0.304 & 0.271 &  0.062 &  0.016 & 0.382 &II-2\\
\hline
\end{tabular}
\end{table}


\begin{figure}
\centerline{\epsfig{file=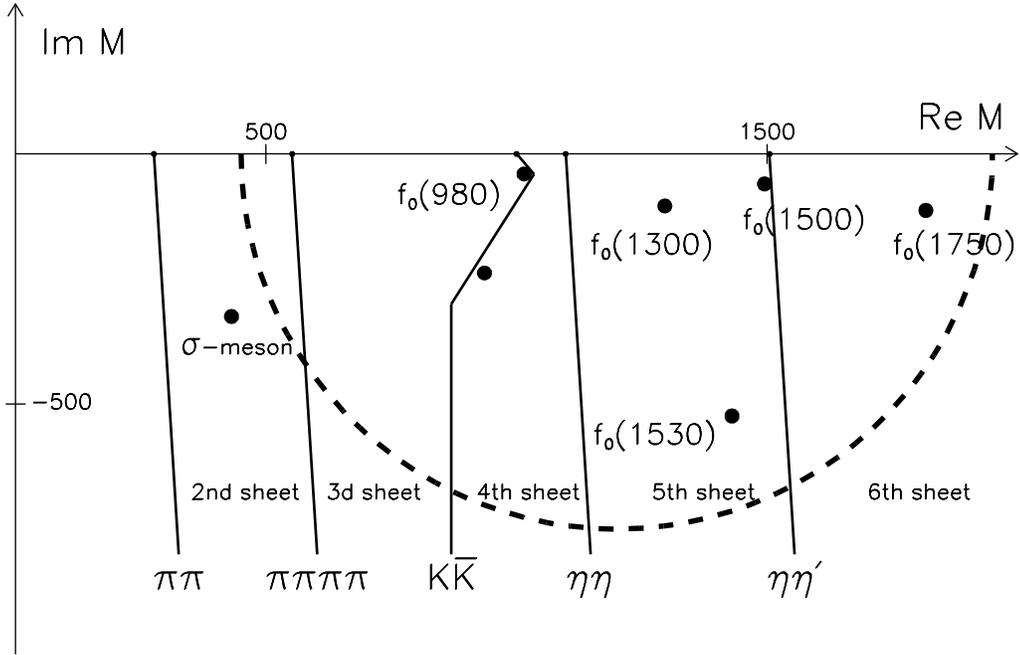,width=14cm}}
\caption{Pole positions (full circles) in the complex-$M$ plane
($M=\sqrt{s}$). Solid lines stand for cuts related to the threshold
singularities ($\pi\pi$, $\pi\pi\pi\pi$, $K\bar K$, $\eta\eta$ and
$\eta\eta'$). Two poles, which correspond to $f_0(980)$, are shown: on
the 3rd and 4th sheets. On the 5th sheet the poles for $f_0(1300)$,
$f_0(1500)$ and broad state $f_0(1530^{+90}_{-250})$ are located
(for the broad state the pole stands at $(1430-i525)$ MeV, that
is the mass for Solution II of the fit [1]).
On the 6th sheet there is a pole for $f_0(1750)$.
The dashed semi-circle restricts
the area where the $K$-matrix fit [1],
which was carried out on the real axis
in the interval $450\;{\rm MeV} \le M \le 1900\;{\rm MeV} $,
can give a reliable reconstruction of analytical amplitude.}
\end{figure}

\begin{figure}
\centerline{\epsfig{file=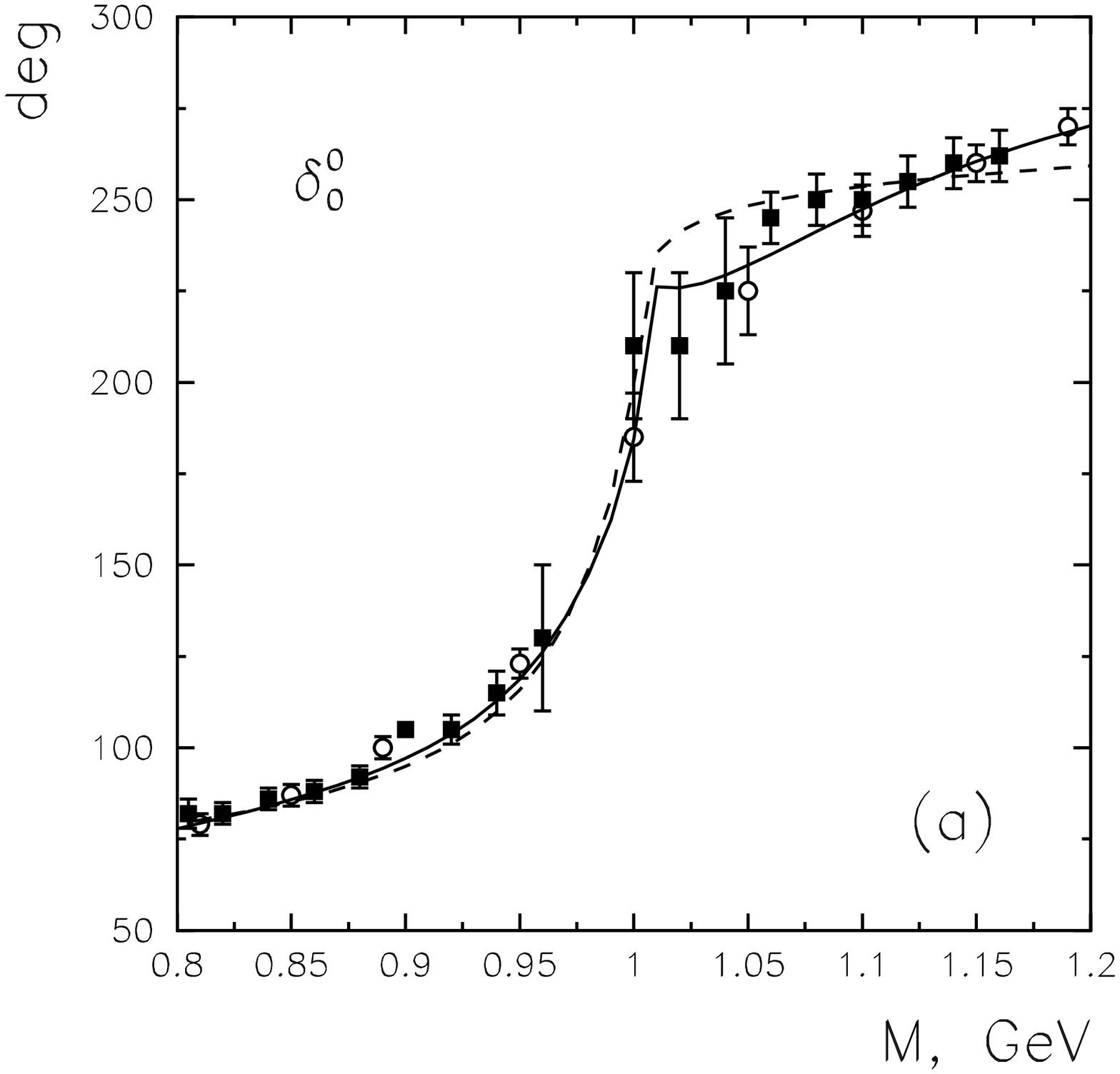,width=7.5cm}
            \epsfig{file=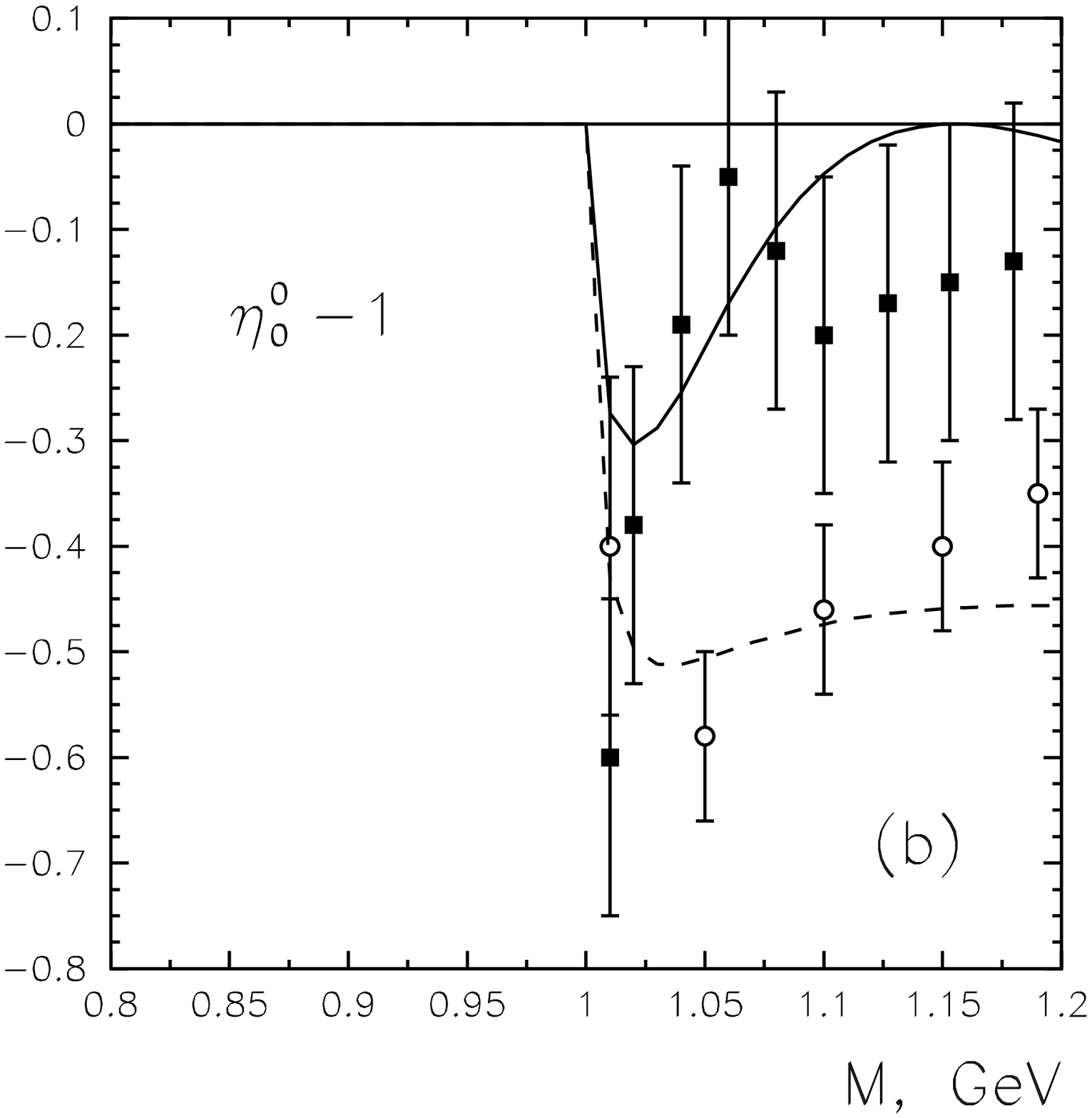,width=7.5cm}}
\caption{Reaction $\pi\pi\to\pi\pi$: description of $\delta^0_0$ and
$\eta^0_0$ in the region of $f_0(980)$. Solid and dashed curves
correspond to the parameter sets A and B. Data are taken from [9] (full
squares) and [10] (open circles). }
\end{figure}

\begin{figure}
\centerline{\epsfig{file=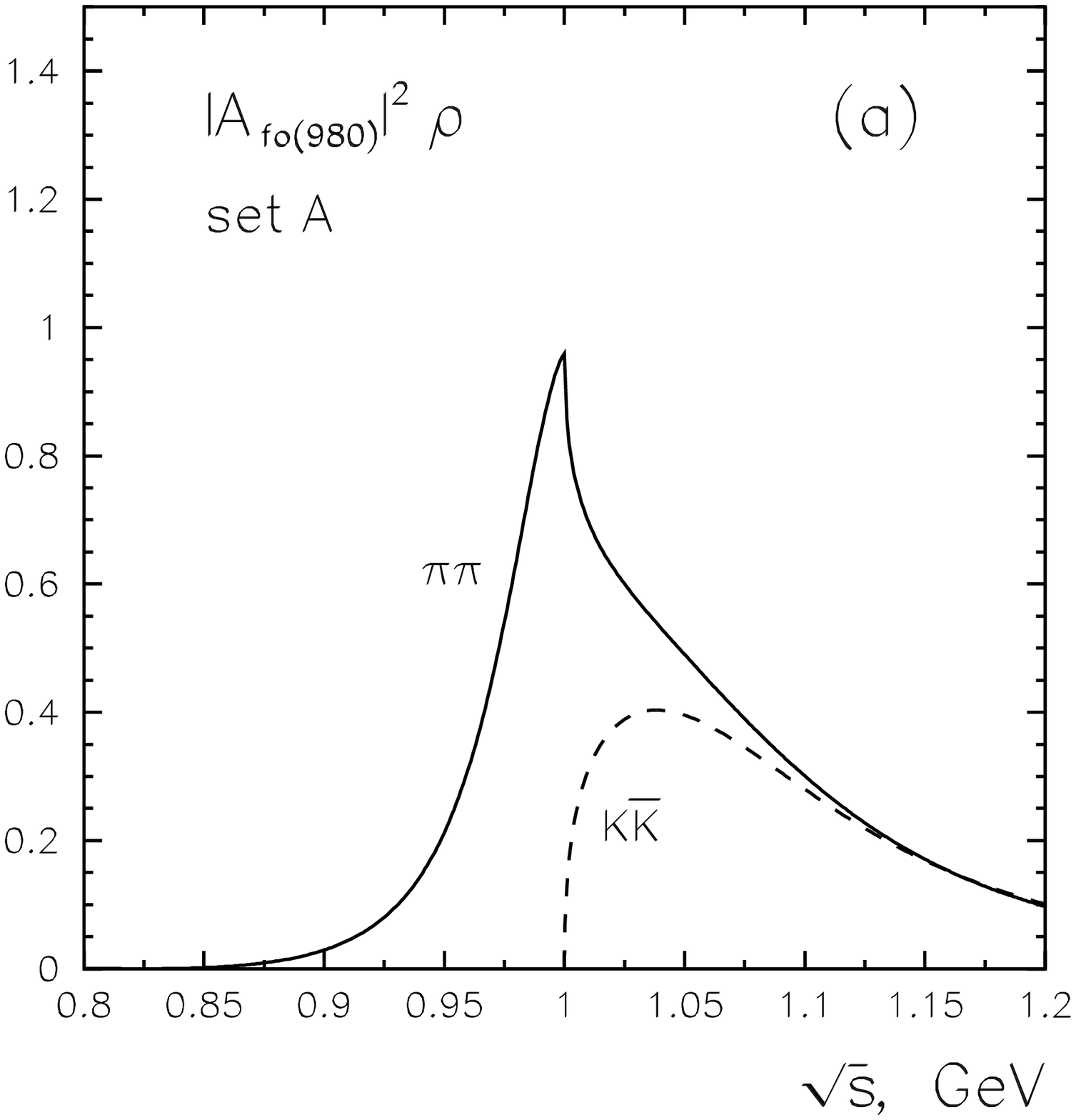,width=7.5cm}
            \epsfig{file=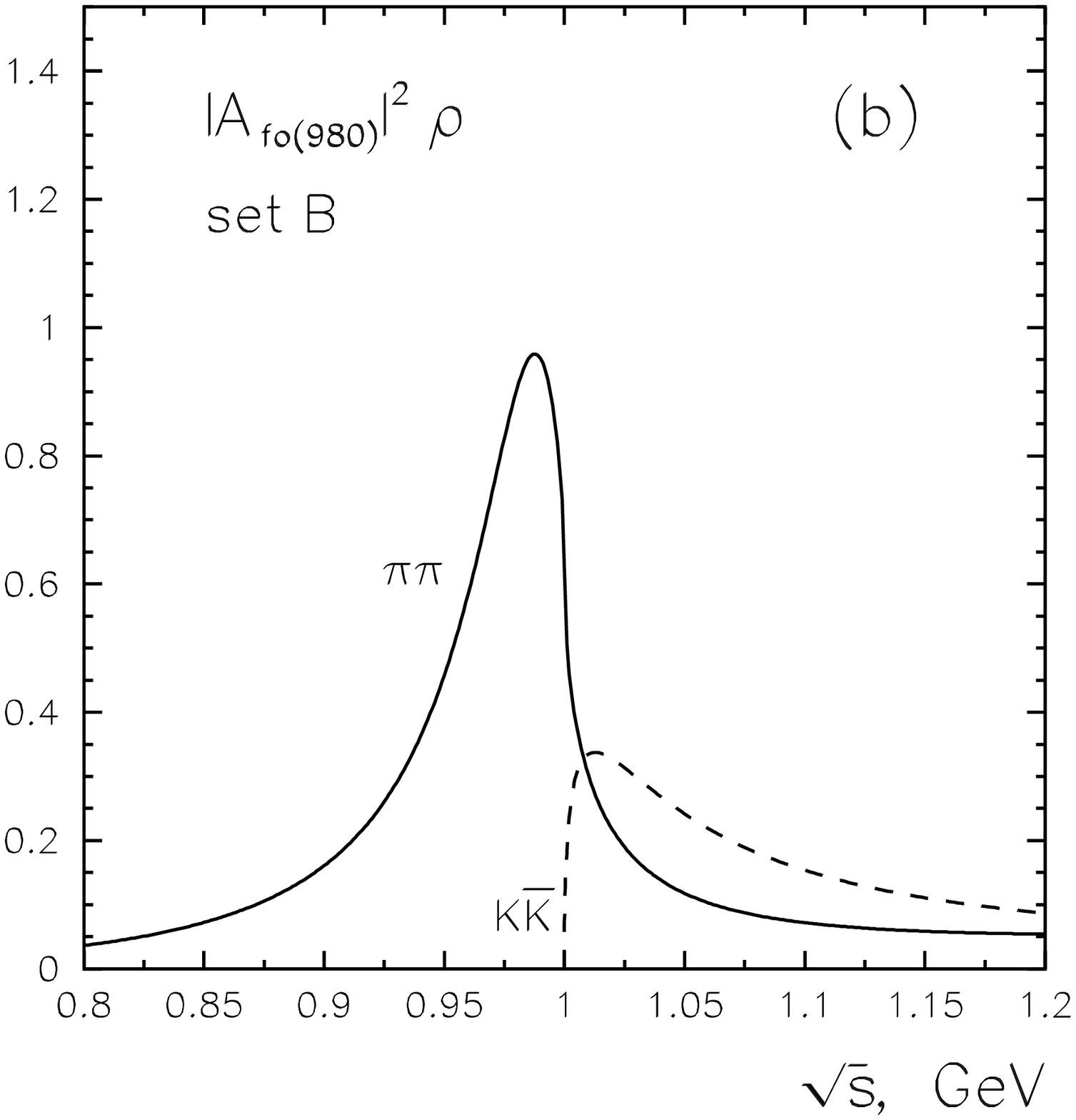,width=7.5cm}}
\centerline{\epsfig{file=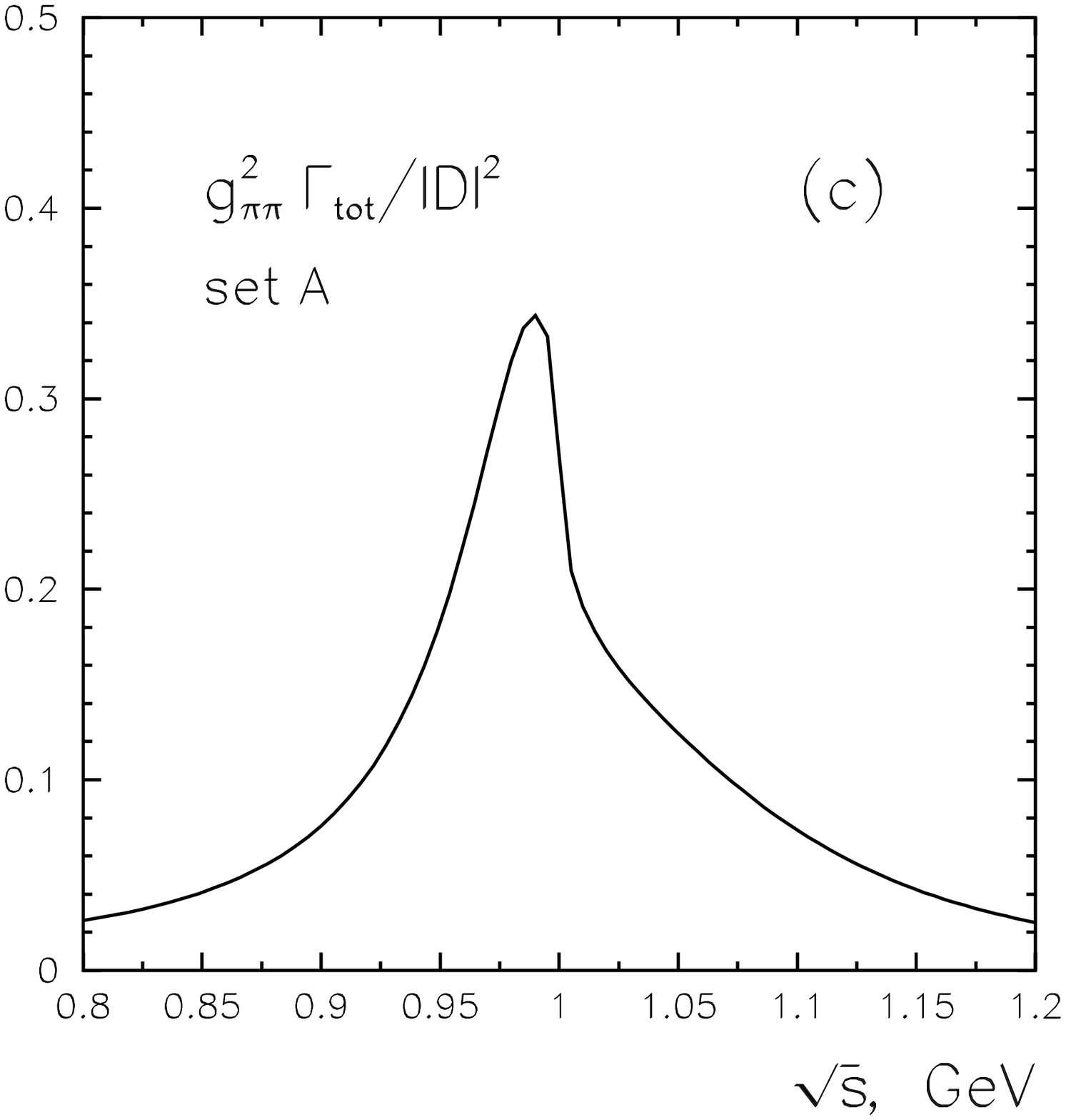   ,width=7.5cm}
            \epsfig{file=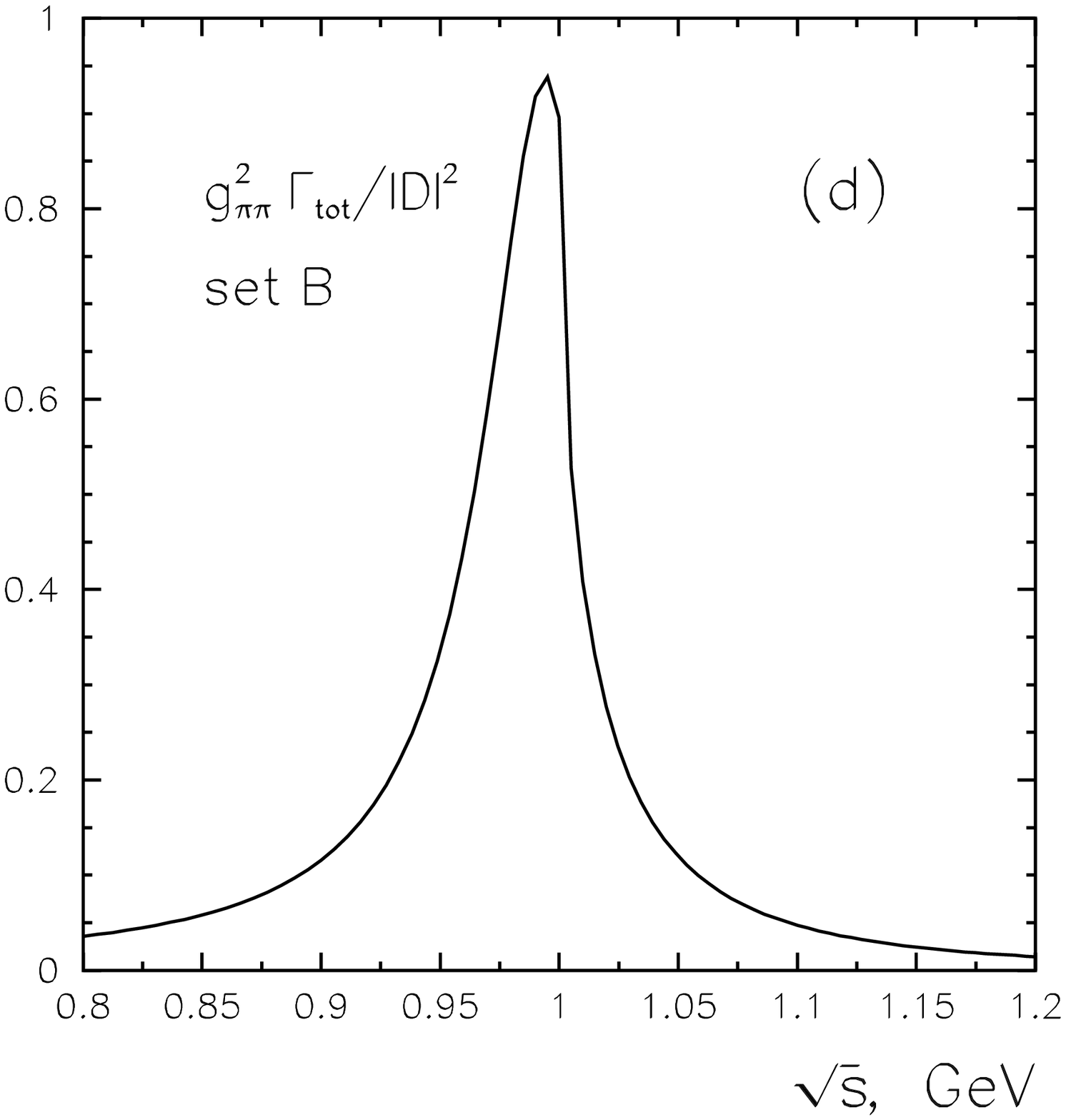   ,width=7.5cm}}
\caption{a,b) The magnitudes
$|R^{(\pi\pi, \pi\pi)}_{f_0(980)}|^2\rho_\pi (s)$ (solid
curve) and  $|R^{(K\bar K, K\bar K)}_{f_0(980)}|^2\rho_K (s)$
(dashed curve) for
the parameter sets A and B. Visible peaks in the $\pi\pi$ spectra
have the total widths $\sim$ 60 MeV (set A) and $\sim$ 45 MeV (set B).
c,d) Values $g_{\pi\pi}^2 \Gamma_{tot}/|D(s)|^2$ (where $g_{\pi\pi}^2
\Gamma_{tot}=$ 0.006 GeV$^2$) for the parameter sets A and B.
Visible total widths of the peaks are $\sim$ 55 MeV (set A) and $\sim$
45 MeV (set B).}
\end{figure}

\end{document}